\documentclass[runningheads]{llncs}
\usepackage[T1]{fontenc}
\usepackage{enumitem}
\setitemize{noitemsep,topsep=0pt,parsep=0pt,partopsep=0pt}
\usepackage{graphicx}
\usepackage{url}
\usepackage{xcolor}

\newif\ifanonym
\anonymfalse

\begin{document}
\title{Shaping Credibility Judgments in Human–GenAI Partnership via Weaker LLMs: A Transactive Memory Perspective on AI Literacy}
\titlerunning{Credibility Calibration in Human-GenAI Partnership via Weaker LLMs}

%
%
\ifanonym
  \author{Anonymous Author(s)}
  \authorrunning{Anonymous}
  \institute{}
\else
  \author{Md Touhidul Islam\orcidID{0000-0002-6075-2832} \and
  Mahir Akgun\orcidID{0000-0001-6884-0119} \and
  Syed Billah\orcidID{0000-0001-5063-3808}}
  \authorrunning{Touhidul, Mahir, and Billah}
  %
  \institute{The Pennsylvania State University, University Park  PA 16802, USA 
  }
\fi
\maketitle              
\begin{abstract}
Generative AI (GenAI) is increasingly being used as a ``knowledge partner'' in higher education. As a result, it increases the need for instructional designs that emphasize GenAI literacy practices such as evaluating the credibility of AI output and maintaining human accountability. 
Existing AI literacy frameworks focus more on what learners should do, and less on how these practices are enacted during routine student-GenAI collaboration. We address this gap by framing student-GenAI interaction as a transactive memory partnership, where credibility is treated as a partnership process that regulates reliance and verification.
To make credibility work visible during coursework, we instantiated GenAI using a weaker, large language model (LLM), that is small enough to run on most students’ computers during in-class activities, yet powerful enough to be helpful, though not so powerful that it eliminates the need for human verification.
In an undergraduate STEM course, we randomly assigned students to one of three instructional conditions that varied how this weaker LLM could be used in repeated activities: reflection-first (think first, then consult AI), verification-required (use AI, then evaluate the output), and control (unrestricted use). Students completed a transactive memory survey at three time points ($N = 42$). Weighted credibility diverged by condition over time. ANCOVA controlling for baseline credibility showed a condition effect at mid-semester, $F(2, 38) = 4.02$, $p = .026$, $\eta_p^2 = .175$, and a stronger effect at post-intervention, $F(2, 38) = 5.48$, $p = .008$, $\eta_p^2 = .224$; adjusted means were lowest in reflection-first, intermediate in verification-required, and highest in control. Parallel analyses for specialization-coordination measures were not significant.
These findings suggest that workflow sequencing, deliberate use of weaker LLMs, and accountability cues embedded in assignment instructions, can recalibrate credibility judgments in students’ GenAI usage, with reflection-first producing the strongest downward shift in reliance.

\end{abstract}

\section{Introduction}

Generative artificial intelligence (GenAI) systems have become a routine presence in students’ academic work~\cite{DEC2024students}. Students use these systems for ideation, explanation, critique, and decision-making, and these interactions increasingly resemble collaboration with a ``knowledge partner.'' However, this shift raises a key challenge for students. They must decide when AI output is believable, relevant, and appropriate for the task at hand, as well as how much responsibility they retain when they incorporate AI contributions into their work. These credibility judgments affect students' learning processes and academic outcomes in several ways. For instance, they influence whether students accept flawed output, over-correct useful suggestions, or disengage from productive support. Recent AI literacy frameworks place credibility evaluation and human accountability at the core of responsible AI use in education~\cite{long2020ai}.

AI literacy frameworks articulate practices that support responsible use, including evaluating outputs, recognizing uncertainty and limitations, and maintaining human accountability when creating with AI~\cite{almatrafi2024systematic,ng2022using}. These accounts clarify what learners should do in principle. However, coursework requires additional clarity about how learners carry out credibility work during repeated human-GenAI collaboration. Since students engage in cycles of delegation, evaluation, revision, and verification, instructional contexts can shape these cycles in systematic ways. Reviews of AI-integrated learning highlight a shortage of empirical evidence that links assignment-level design choices to students’ credibility judgments over time, especially when AI is embedded across recurring course activities~\cite{van2025landscape}.

In this paper, we use Transactive Memory Systems (TMS) theory as a process-oriented lens for specifying these mechanisms~\cite{wegner1987transactive}. TMS theory explains how partners distribute cognitive work through beliefs about where knowledge resides, how credible it is, and how contributions are coordinated during task execution~\cite{wegner1987transactive,lewis2003measuring}. TMS has three dimensions: \emph{\textbf{Specialization}} captures perceived division of expertise; \emph{\textbf{Credibility}} captures perceived reliability of a partner’s knowledge; and \emph{\textbf{Coordination}} captures how smoothly partners integrate contributions. In AI-mediated learning, students continuously decide what to delegate to GenAI, how trustworthy its outputs appear, and how to integrate those outputs with their own reasoning and disciplinary standards. From this perspective, credibility evaluation practices, which are emphasized in AI literacy, regulate the partnership during student-GenAI collaboration.

To apply TMS theory in practice, \textbf{\textit{we propose instantiating GenAI using a weaker large language model (LLM)}}, such as Qwen~2.5:7B~\cite{hui2024qwen2}, which students can run on their own computers during in-class activities. These models support ideation and explanation; however, they have shorter context windows and longer response times. As such, students must write precise prompts, limit the number of turns, and test outputs against course materials. As activities were repeated, the weaker LLM created regular opportunities for scrutiny and recalibration that are harder to elicit when a commercial system (e.g., ChatGPT) produces fluent, confident responses that can inflate perceived certainty.

We also propose assignment-level workflow sequencing to study the effects of the three dimensions of TMS (credibility, specialization, and coordination) over time. In an undergraduate STEM course, we randomly assigned students to one of \textbf{three} conditions that varied the required workflow for AI use: i) the \emph{\textbf{reflection-first}} condition, where students generated an initial response before consulting the weaker LLM; ii) the \emph{\textbf{verification-required}} condition, where students used the weaker LLM and then evaluated the output; and iii) the \emph{\textbf{control}} condition that allowed unrestricted use.

Our results show that \textbf{\textit{credibility judgments diverged by condition across the semester, with the strongest downward shift in the reflection-first condition}}. Specialization and coordination measures did not show comparable changes. These findings suggest that sequencing and accountability cues in assignment instructions can recalibrate students’ credibility judgments even when the AI system remains constant.

\textbf{Contributions:}
In sum, we make the following contributions.
\begin{itemize}
    \item \textbf{A process account of GenAI literacy.} We frame credibility evaluation in AI literacy as partnership regulation within a transactive memory system, which specifies how reliance and verification can be shaped by instructional context.
    \item \textbf{A design rationale for weaker LLMs in coursework.} We show how a locally runnable weaker LLM can make credibility work observable during in-class activities and can support instructional interventions that target verification routines.
    \item \textbf{Empirical evidence from a randomized field study.} We report how three workflow conditions influence credibility judgments over time, using repeated measures of TMS dimensions in an undergraduate course.
    
\end{itemize}

\section{Background and Related Work}
AI literacy is widely recognized as a multidimensional construct that includes knowledge, skills, and dispositions for responsible engagement with AI. Systematic reviews and frameworks substantiate this framing across educational and psychological research~\cite{almatrafi2024systematic,ng2022using,long2020ai,mills2024ai,OECD2025}. Across contexts, these frameworks emphasize learners’ ability to evaluate AI output, recognize limitations and uncertainty, and create with AI while retaining human accountability for judgment and decision-making~\cite{mills2024ai,OECD2025}. This breadth supports adoption across tasks and disciplines, and it also means that specific competencies can be enacted through different routines in everyday coursework.

In this study, we focus on critical thinking and evaluation as a core AI literacy dimension. We treat this dimension as a disposition that develops through repeated engagement with AI in context, where students’ practices are shaped by task demands and instructional structures. To specify mechanisms that connect instructional routines to credibility work in student-GenAI interaction, we draw on Transactive Memory Systems (TMS) theory as a process-oriented account of how partnerships organize delegation, evaluation, and responsibility~\cite{wegner1987transactive}.

\subsection{Transactive Memory Systems as Partnership Processes}
TMS theory describes how cognitive work is distributed across a system of agents through shared expectations about knowledge location, access, and reliability~\cite{wegner1987transactive}. The unit of analysis is the partnership process that governs when an agent retrieves information externally, when an agent relies on it, and when an agent invests effort in verification and integration. Empirical work commonly operationalizes these processes through specialization, credibility, and coordination~\cite{lewis2003measuring,peltokorpi2008transactive}. In this paper, these dimensions serve as vocabulary for describing how a student-GenAI partnership functions over time, with credibility as the focal process that regulates reliance and verification during task execution.

TMS has an established measurement tradition that supports educational field studies. The three-dimensional structure has been used to characterize collaboration quality across domains, and it supports longitudinal designs that examine how partnership beliefs shift with experience and feedback~\cite{lewis2003measuring}. This orientation aligns with coursework settings where students repeatedly revisit similar activity structures and gradually develop stable patterns of delegation and checking.

\subsection{TMS for Human-GenAI Collaboration}
TMS was originally developed for human teams, and its mechanisms generalize to contexts where an external resource supplies information that can be retrieved, evaluated, and integrated into ongoing work. GenAI systems fit this description, since they provide candidate knowledge and reasoning traces that students can incorporate into problem solving. The partnership challenge becomes calibration. Students must decide when GenAI output merits acceptance, when it requires verification, and how to reconcile AI-generated content with disciplinary standards and personal reasoning.

Reliability profiles of AI systems shape these partnership processes. A weaker LLM that produces variable and error-prone outputs increases the salience of credibility regulation during routine task execution. Students encounter omissions, inconsistencies, and context breakdowns frequently enough that verification becomes a recurring step, and coordination becomes visible through how students sequence prompting, checking, and revision. This framing supports a process account that distinguishes task-level credibility regulation from global trust attitudes, and it provides a way to interpret how instructional cues and workflow constraints can shift reliance patterns over time.

\subsection{Weaker, Locally Runnable LLMs in Classroom Settings}
Course deployments frequently rely on cloud-hosted assistants that deliver high fluency, long contexts, and low latency. Locally runnable weaker LLMs shift the interaction to a constrained setting that students can execute and inspect on personal machines, and this shift changes what credibility work looks like during coursework. Model size and deployment choices jointly determine feasibility and limitations. A 7B-parameter model (e.g., Qwen 2.5:7B~\cite{hui2024qwen2}) requires roughly 14 GB of weights in FP16, and quantization to 8-bit or 4-bit reduces this footprint to roughly 7 GB or 3-4 GB, which supports execution on many laptops. Context length matters as much as parameter count, since each additional token increases the key-value cache used during autoregressive generation, and longer multi-turn dialogues increase memory use and response time. Students therefore work under tighter context budgets and turn budgets, and these constraints shape how they prompt, how they manage information, and how they verify claims against course materials. Weaker models also exhibit more variable accuracy and shorter-range coherence, and quantization can amplify this variability in ways that make breakdowns easier to notice. Cloud models (e.g., ChatGPT, Claude) typically provide stronger instruction following, broader knowledge coverage, and longer effective contexts, and their fluent responses can elevate perceived certainty. In our setting, the local weaker LLM provides sufficient utility for ideation and explanation, and it preserves a reliability profile that keeps verification and accountability salient across repeated activities.

\subsection{Connecting AI Literacy and Transactive Memory Systems}
AI literacy frameworks specify competencies and dispositions that support responsible AI use; TMS specifies partnership processes that organize those competencies during collaboration. In this study, we use TMS to characterize the partnership processes through which critical evaluation is enacted during repeated student-GenAI work. This connection positions workflow sequencing and accountability cues as levers that can shape credibility regulation in situ, with specialization and coordination as complementary indicators of how the partnership is structured and enacted over time.

\subsection{Present Study}
In this study, we focus on credibility judgments as a transactive memory process that aligns with the critical thinking and evaluation dimension of AI literacy. While TMS encompasses specialization and coordination alongside credibility, the present study targets credibility because of its close theoretical connection to evaluative dispositions emphasized in AI literacy frameworks. Other AI literacy dimensions and partnership processes remain important but are beyond the scope of the present investigation.

\section{Methods}

\subsection{Study Design}

We used a between-subjects experimental design with three instructional conditions and three waves of measurement (pre-intervention, mid-course, and post-intervention). Students were randomly assigned to one of three conditions that differed in the required workflow for AI use during repeated in-class activities:

\begin{itemize}
    \item \textbf{Reflection-first}: Students began each activity by reasoning through the task independently and articulating an initial approach or solution before consulting the AI system. AI use served as a secondary resource for checking, refining, or extending students’ reasoning, and students began from their own solution path.
    \item \textbf{Verification-required}: Students could consult the AI system at any point during the activity, and they were required to evaluate the AI’s output using course materials and in-class resources (e.g., lecture notes, provided examples, and task-specific datasets). This evaluation included identifying useful and problematic aspects of the output and justifying whether AI-generated suggestions were incorporated into the final work, as well as how they were incorporated.
    \item \textbf{Control}: Students had unrestricted access to the AI system during the activity. The activity included no required reflection on, or evaluation of, AI output.
\end{itemize}

All students completed identical tasks; only the AI-use instructions differed. This design supports attribution of differences in credibility judgments to instructional condition, with task demands held constant.

\subsection{Participants and Procedure}

Participants were undergraduate students enrolled in a visual analytics course at a large public R1 university. A total of $N = 42$ students completed all three survey waves and were included in the longitudinal quantitative analyses, with approximately 12-18 students per condition. Open-ended survey responses were collected at post-intervention and were used for qualitative analysis.

Students completed the same survey instrument at three time points: pre-intervention (Time~1), mid-course (Time~2), and post-intervention (Time~3). Between survey administrations, students engaged in multiple in-class activities involving data visualization tasks and followed the AI-use workflow specified by their assigned condition.

Students used a local 7B-parameter language model (Qwen~2.5 Coder via Ollama) as their AI partner. We selected a constrained local model to achieve a favorable trade-off between utility and visible limitations. In practice, the model supported ideation and explanation during short exchanges, and it exhibited limitations that are salient in local deployment (e.g., smaller effective context, slower responses than cloud models, and a higher propensity for hallucination in longer prompts). These properties created recurring opportunities for students to encounter and evaluate AI fallibility. This design aligns with our transactive memory framing, since it foregrounds credibility judgments as an ongoing partnership process and keeps reliability and verification salient during routine coursework.

At post-intervention, students also responded to open-ended survey items about their experiences working with AI.

\subsection{Measures}

\subsubsection{Quantitative Measures}

We adapted a Transactive Memory Systems (TMS) questionnaire from~\cite{Akgun2026Cultural} and computed factor-weighted scores using the factor loadings. This procedure yielded two factor-weighted constructs:

\begin{itemize}
    \item \textbf{CR-S} (weighted credibility): capturing students’ credibility judgments regarding AI expertise, including willingness to rely on AI output and perceived need for verification.
    \item \textbf{SP-COR} (specialization-coordination): reflecting perceived division of labor and responsibility between human and AI.
\end{itemize}

Weighted credibility (CR-S) served as the primary outcome, since the instructional conditions were designed to shape how students evaluate and calibrate reliance on AI output. Specialization-coordination (SP-COR) served as a secondary outcome and allowed us to examine whether instructional effects concentrated on credibility-related partnership processes or extended to broader aspects of the transactive system.

\subsubsection{Qualitative Measures}
Open-ended survey items administered at post-intervention asked students to describe (1) how their approach to questioning AI output changed over the course, (2) their perceptions of human and AI roles in collaborative tasks, and (3) how the course structure influenced their AI use.

\subsection{Analytic Approach}

\subsubsection{Quantitative Analysis}
To examine whether instructional condition influenced credibility judgments while accounting for baseline individual differences, we conducted analyses of covariance (ANCOVA) using Type III sums of squares. Instructional condition served as a between-subjects factor, and baseline scores (Time~1) served as covariates. We report effect sizes using partial eta squared ($\eta_p^2$).

We evaluated key assumptions prior to analysis. Baseline weighted credibility was comparable across conditions, $F(2, 39) = 1.23$, $p = .30$, which supports its use as a covariate. The homogeneity of regression slopes assumption was met, $F(2, 36) = 0.06$, $p = .94$. Levene’s test indicated comparable variances for Time~3 credibility across conditions, $F(2, 39) = 1.91$, $p = .16$.

\subsubsection{Qualitative Analysis}

Open-ended survey responses were analyzed using thematic analysis~\cite{braun2006using}. Initial codes were generated inductively from the data to capture students’ descriptions of AI interactions, credibility judgments, and collaborative practices. We then organized codes under two higher-order dimensions, credibility and specialization-coordination, to align interpretation with the transactive memory framework. We selected representative quotes to illustrate each theme and attended to coverage across all three instructional conditions.

\section{Findings}

Findings are organized around two transactive memory dimensions: (1) credibility: students' trust in and critical evaluation of AI-generated knowledge, and (2) specialization and coordination: how students and AI divided cognitive labor and organized their collaborative work. For each dimension, we present quantitative results followed by qualitative elaboration. We also report an emergent theme that cuts across dimensions: how the constrained local model functioned as a pedagogical scaffold.

\subsection{Credibility Calibration}

\subsubsection{Quantitative Results}
\begin{figure}[t!]
\centering
\includegraphics[width=.95\linewidth]{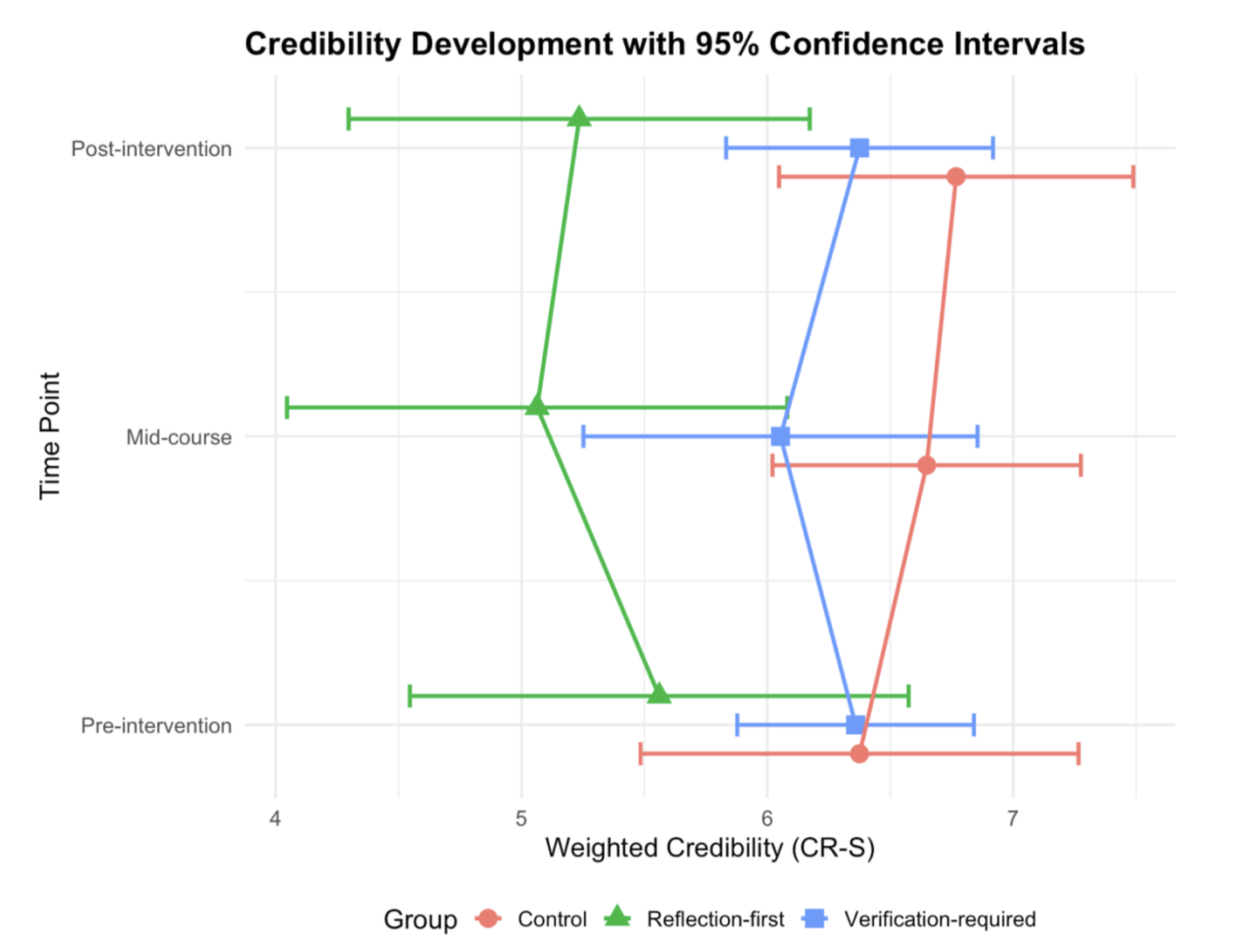}
\caption{The progress of weighted credibility (CR-S) across the course by instructional condition. Points indicate group means; horizontal bars represent 95\% confidence.}
\label{fig:credibility}
\end{figure}

Figure~\ref{fig:credibility} shows trajectories of weighted credibility (\textbf{CR-S}) across three measurement points by instructional condition. At pre-intervention, credibility scores were comparable across conditions with substantial overlap in confidence intervals. By mid-course, group means diverged. Reflection-first and verification-required shifted downward relative to baseline, and control shifted upward. ANCOVA controlling for baseline credibility indicated a significant condition effect at mid-course, $F(2, 38) = 4.02$, $p = .026$, $\eta_p^2 = .175$.

By post-intervention, divergence was larger. ANCOVA controlling for baseline credibility again showed a significant effect of instructional condition, $F(2, 38) = 5.48$, $p = .008$, $\eta_p^2 = .224$. Estimated marginal means followed a consistent ordering: reflection-first (adjusted $M = 5.65$), verification-required (adjusted $M = 6.22$), and control (adjusted $M = 6.60$). Holm-adjusted pairwise comparisons separated reflection-first from control ($p = .006$); the remaining contrasts were smaller and did not reach statistical significance.

\subsubsection{Qualitative Elaboration: How Credibility Calibration Developed}

Qualitative responses clarify how credibility calibration emerged across the semester. Students across all conditions described learning from direct encounters with model errors. Many reported initial acceptance of fluent responses, followed by more cautious evaluation after experiencing hallucinations, omissions, and task-specific inaccuracies. A control student described this shift:

\begin{quote}
\textit{``Throughout this course, my approach to questioning AI has become more careful. At first, I accepted responses without much thought because they looked clear and confident. Over time, I realized that AI can sound convincing even when it is wrong, so I started treating it as something to question and verify rather than simply trust.''}
\end{quote}

Students in the verification-required condition made a similar point while generalizing beyond the local model: \textit{``I double check more. In the class I saw that if small models can hallucinate to an extent that it changes the final output, I can't imagine how much bigger models hallucinate.''} 

Similarly, students in the reflection-first condition stated routine cross-checking: \textit{``Throughout this class, I have started to double-check what the AI is saying because there have been multiple instances where it has been on the right track, but not all of its responses were 100\% accurate.''}

Students’ accounts also show condition-specific pathways that align with the quantitative ordering.

\paragraph{
\textbf{Reflection-first.}
}
Students described the workflow as anchoring activity execution in human reasoning, with AI positioned as a secondary resource. This framing reduced credibility attributions by design, since students arrived at an initial answer before consulting the model:
\textit{``Because the course required me to think first and then verify AI output, it made me treat AI as a tool to double-check ideas rather than something to trust automatically.''}

Another student emphasized sequencing and accountability in terms of when AI input was admissible: \textit{``I trusted AI to verify my work or improve it based on its suggestions. I most definitely did not trust it to first provide me a baseline to work off of to complete a task---I need to completely understand the task and expected output, complete a draft, and then the AI can offer suggestions.''} 

A third student described a forward transfer expectation: \textit{``Sometimes, prior to this class, I would just ask AI for the answer to a question before doing anything else. After doing these in-class activities, I will definitely try to think of my own ideas before consulting AI in the future.''}

\paragraph{
\textbf{Verification-required.}
}
Students described developing evaluation routines through mandated checking and justification. These routines often centered on claim-by-claim inspection against course materials:
\textit{``There were a few instances where AI failed to accurately identify key issues or make key suggestions that were needed to completely fulfill an understanding. I was able to catch these things, because I went through the AI output with a fine comb to verify the claims it had made.''}

Another student framed the requirement as shaping attention and strategy: \textit{``This made me more critical of AI responses and how to leverage AI to make my work more efficient.''} In the quantitative results, verification-required fell between reflection-first and control, which suggests that required verification supported critical habits while reflection-first produced the largest shift in credibility judgments.

\paragraph{
\textbf{Control.}
}
Control students reported heterogeneous strategies. Some developed skeptical routines through experience and feedback: \textit{``I think this helped me learn the proper way to use AI. Not only did it show me how I should work with AI as a tool to help not to fully complete things for me, but it showed me how to think when using AI.''} 

Others retained higher trust when inputs were well-scoped: \textit{``I trusted the AI a bit more when it came to activities because we provided the AI the data needed when asking questions catered to the activity.''} One student explicitly noted over-reliance: \textit{``I think that Ollama hindered my learning a bit because I tended to rely overly on the LLM instead of using my own knowledge.''} These accounts align with the quantitative pattern in which control showed the highest post-intervention credibility.

\subsection{Specialization and Coordination}

\subsubsection{Quantitative Results}
ANCOVA for the specialization-coordination composite (SP-COR) showed no evidence of condition differences. This pattern suggests that the instructional manipulations concentrated on evaluative partnership processes and left broader divisions of cognitive labor relatively stable at the scale captured by SP-COR. Qualitative responses still revealed meaningful variation in how students described role division and workflow organization.

\subsubsection{Qualitative Elaboration: Role Specialization}
Students frequently articulated distinct domains of expertise for themselves and for AI, which reflects role differentiation consistent with transactive memory theory. Three recurring role configurations emerged.

\paragraph{
\textbf{Human as Decision-Maker, AI as Executor.}
}
Many students described themselves as directing strategy and validating outcomes while delegating implementation steps, as mentioned by a student in Verification-required condition:
\textit{``I see my role as a final decision maker and a manager while the AI is a very efficient but sometimes incorrect employee of mine.''}

A reflection-first student similarly described directional control: \textit{``I see myself as the main generator of direction and decision making where as the AI is the executor of basic actions to complete the task.''}

\paragraph{
\textbf{Human as Evaluator, AI as Generator.}
}
Other students cast AI as producing candidate content and themselves as verifying correctness:
\textit{``AI is good for generation of content, while I am good at verifying.''} (Verification-required)

One student summarized this relationship through assessment roles: \textit{``I am an exam proctor/grader and it is an exam taker.''} (Verification-required)

\paragraph{
\textbf{Collaborative Partnership.}
}
Some students described a more symmetric collaboration:
\textit{``I think we are at the same level. AI and I help each other and work together to complete tasks.''} (Control)

Another student characterized the relationship as \textit{``coordinating with a partner that is wiser than you.''} (Reflection-first)

These configurations appeared across all conditions, which is consistent with the non-significant quantitative differences on SP-COR.

\subsubsection{Qualitative Elaboration: Coordination Through Prompting}
Students described coordination as a communication problem that required learning how to specify constraints, context, and intent through prompts. Early prompts were often underspecified:
\begin{quote}
    
\textit{``I learned a lot about how to prompt the AI to get the information that I need. I used to just give it commands, but now I'm better at providing background information so the AI can give the best output possible.''} (Control)
\end{quote}
Students connected low-quality output to missing context: \textit{``Sometimes when I just copy paste the question to the AI, but did not give it the background of the question, I will not trust the AI, because it always gives the answers based on what it receives.''} (Control) 
Over time, students reported improved coordination through more specific prompting: \textit{``I learned to be more descriptive when giving prompts which lead to less messages after the fact clarifying what I meant.''}

\subsection{The Constrained Model as Pedagogical Scaffold}

An emerging theme was that the local model's limitations functioned as productive constraints. Students explicitly contrasted their experience with commercial alternatives:
\textit{``Local LLMs are better for educational purposes since it isn't able to give you complex answers, requiring you to think on your own.''} (Control)

Several students described how constraints prompted deliberate prompting and verification: \textit{``I mistrust AI even more because Ollama gave shaky responses. I did learn how to prompt more effectively, because I knew the LLM was limited.''} (Control) Another student contrasted interactional style: \textit{``The local LLM does not have the `agreeableness' and `yes-man' quality that ChatGPT has, which was refreshing.''} (Control) 

A verification-required student emphasized transfer to future AI use: \textit{``It helped my learning as I learned how to better use LLMs, and it has taught me to genuinely question everything, as having blind trust is not a great way to work when using LLMs such as this.''}

\section{Discussion}
The quantitative and qualitative findings clarify how partnership perceptions relate to AI literacy–relevant evaluation. The survey results indicate that students’ credibility judgments about GenAI shifted over time, reflecting changes in how reliable they perceived the AI partner to be and how readily they were willing to rely on it. Students’ written responses provide meaning for these shifts: lower credibility was not simply ``less liking'' of AI, but a stronger stance of active judgment—describing habits such as thinking independently first, treating AI as a source to be checked, and verifying outputs against course standards. In this sense, the quantitative trends capture changes in perceived partnership credibility, while the qualitative accounts show how those changes are experienced and articulated as critical evaluation practices, connecting a TMS process (credibility regulation) to the critical thinking and evaluation dimension of AI literacy.

\paragraph{\textbf{Implications for Transactive Memory Systems (TMS) and AI Literacy}}
This study extends TMS theory to student–GenAI collaboration by showing that credibility judgments can be shaped by how course activities structure students’ engagement with AI. In TMS terms, credibility operates as an ongoing form of judgment that regulates when learners rely on a partner and when they verify ~\cite{wegner1987transactive,lewis2003measuring}. Our findings suggest that relatively small differences in course expectations, such as whether students must form an initial answer before consulting AI, or must justify an evaluation of AI output, can shift how students judge the reliability of AI over time. At the same time, the absence of comparable changes in specialization–coordination indicates that instruction can influence credibility without necessarily changing how students divide roles with AI. This distinction reinforces the idea that TMS comprises multiple partnership processes that do not always progress in tandem, and that different instructional supports may target different aspects of the partnership.

These results also strengthen the connection between TMS and AI literacy. AI literacy frameworks identify critical thinking and evaluation as central~\cite{mills2024ai,OECD2025}, but they often describe these practices at the level of what learners should do. Our results suggest a complementary account: critical evaluation can be supported by shaping how learners regulate credibility in collaboration with AI, particularly when course structures discourage default reliance and promote active judgment. In this way, TMS helps explain how instructional design can support an AI literacy–relevant evaluative stance without treating AI literacy as a single measured outcome.

\paragraph{\textbf{Weaker Local Models as Pedagogical Scaffolds Through Epistemic Friction}}
Students’ reflections highlight the local weaker LLM as an instructional affordance, which can be a design principle for AI-integrated coursework. A weaker local model introduces friction that can support learning goals, as long as the friction stays within a productive range. 
In contrast, highly capable AI models can compress interaction into a smooth pipeline that hides uncertainty and reduces opportunities for epistemic checking. 
However, students can experience frustration when latency and context constraints dominate the activity. Instruction can manage this trade-off by constraining the scope of AI use, providing prompt templates, limiting the number of turns, and aligning tasks so that AI output must be reconciled with course materials. This framing also broadens the educational conversation about model choice.

\paragraph{\textbf{Future Work}}
This study opens several avenues for further investigation. One natural extension involves developing behavioral measures of credibility assessment. Prompt logs, revision histories, and artifact comparisons could operationalize verification as observable behavior, making it possible to identify when students cross-check AI output, when they accept it verbatim, and when they integrate suggestions through substantive transformation.

A second direction concerns generalizability and persistence. Replication across disciplines, course levels, and institution types would test whether these findings hold in broader contexts, while follow-up measures in subsequent courses could assess whether reflection-first habits endure and transfer to more capable cloud-based systems.

Finally, model capability itself warrants treatment as a continuous design variable. Studies that systematically vary model strength, context length, quantization, and latency could help characterize the thresholds at which epistemic friction shifts from productive challenge to undue burden.

\paragraph{\textbf{Limitations}}
Several limitations bound interpretation. First, the sample size was modest ($N = 42$) and drawn from a single course at a single institution. This limits generalizability and statistical power, especially for detecting smaller effects on specialization and coordination. Second, the primary outcomes were self-reported perceptions measured through factor-weighted TMS constructs. These measures capture partnership beliefs and dispositions, and they do not directly measure verification behavior, prompt content, or learning outcomes in the course. Finally, the intervention was embedded in in-class activities with a specific local model (Qwen 2.5 Coder via Ollama). Other tasks, domains, and model profiles may yield different dynamics.

\section{Conclusion}
This paper positions student--GenAI interaction as a transactive memory partnership, treating credibility assessment as a regulatory process that shapes patterns of reliance and verification during coursework. 
A classroom experiment examined how workflow sequencing and accountability cues influenced credibility calibration over time. 
Results showed that reflection-first conditions produced the strongest downward shift in perceived AI credibility relative to unrestricted use, while specialization-coordination conditions remained comparatively stable. 
Qualitative responses further illuminated how students developed role differentiation and learned to coordinate with the AI through iterative prompting.

These findings carry practical implications for instruction. They suggest that routine workflow design can serve as a lever for fostering critical evaluation skills central to AI literacy, and that weaker local LLMs---despite their limitations---may function as effective pedagogical scaffolds precisely because their constraints keep verification salient.

\bibliographystyle{splncs04}
\bibliography{Bibliography, Bibliography2, Bibliography3}

@article{braun2006using,
	title        = {Using thematic analysis in psychology},
	author       = {Braun, Virginia and Clarke, Victoria},
	year         = 2006,
	journal      = {Qualitative research in psychology},
	publisher    = {Taylor \& Francis},
	volume       = 3,
	number       = 2,
	pages        = {77--101}
}

@string{Computing = "Computing"}

@string{Springer = "Springer-Verlag"}

@misc{chatgpt,
  title = {ChatGPT},
  author = {OpenAI},
  year = {2020},
  howpublished = {\url{https://openai.com/blog/chat-turing-test-6-billion-parameter-language-model/}}
}

@inproceedings{Akgun2026Cultural,
  author    = {Akgun, Mahir and Toker, Sacip},
  title     = {Cultural Variations in Human-{AI} Partnership: Initial Cross-Cultural Validation of the Transactive Memory System with {GenAI} ({TMS}-{GenAI}) Measurement Tool},
  booktitle = {Proceedings of the 2026 CHI Conference on Human Factors in Computing Systems},
  series    = {CHI '26},
  year      = {2026},
  date      = {2026-04},
  location  = {Barcelona, Spain},
  publisher = {Association for Computing Machinery},
  address   = {New York, NY, USA},
  numpages  = {18},
  url       = {https://doi.org/10.1145/3772318.3790980},
  doi       = {10.1145/3772318.3790980}
}

@techreport{OECD2025,
  author      = {{OECD}},
  title       = {Empowering Learners for the Age of AI: An AI Literacy Framework for Primary and Secondary Education (Review Draft)},
  institution = {OECD},
  year        = {2025},
  date        = {2025-05},
  type        = {Review Draft},
  address     = {Paris},
  url         = {https://ailiteracyframework.org},
  note        = {Co-funded by the European Union. With support from Code.org}
}

@article{mills2024ai,
  title={AI Literacy: A Framework to Understand, Evaluate, and Use Emerging Technology.},
  author={Mills, Kelly and Ruiz, Pati and Lee, Keun-woo and Coenraad, Merijke and Fusco, Judi and Roschelle, Jeremy and Weisgrau, Josh},
  journal={Digital Promise},
  year={2024},
  publisher={ERIC}
}

@techreport{DEC2024students,
  author      = {{Digital Education Council}},
  title       = {What students want: Key results from {DEC} {Global} {AI} {Student} {Survey} 2024},
  institution = {Digital Education Council},
  year        = {2024},
  month       = aug,
  day         = {7},
  url         = {https://www.digitaleducationcouncil.com/post/what-students-want-key-results-from-dec-global-ai-student-survey-2024%7D},
  note        = {Accessed: 2026-01-30}
}

@article{almatrafi2024systematic,
  title={A systematic review of AI literacy conceptualization, constructs, and implementation and assessment efforts (2019--2023)},
  author={Almatrafi, Omaima and Johri, Aditya and Lee, Hyuna},
  journal={Computers and Education Open},
  volume={6},
  pages={100173},
  year={2024},
  publisher={Elsevier}
}

@article{ng2022using,
  title={Using digital story writing as a pedagogy to develop AI literacy among primary students},
  author={Ng, Davy Tsz Kit and Luo, Wanying and Chan, Helen Man Yi and Chu, Samuel Kai Wah},
  journal={Computers and Education: Artificial Intelligence},
  volume={3},
  pages={100054},
  year={2022},
  publisher={Elsevier}
}

@inproceedings{long2020ai,
  title={What is AI literacy? Competencies and design considerations},
  author={Long, Duri and Magerko, Brian},
  booktitle={Proceedings of the 2020 CHI conference on human factors in computing systems},
  pages={1--16},
  year={2020}
}

@incollection{wegner1987transactive,
  title={Transactive memory: A contemporary analysis of the group mind},
  author={Wegner, Daniel M},
  booktitle={Theories of group behavior},
  pages={185--208},
  year={1987},
  publisher={Springer}
}

@article{lewis2003measuring,
  title={Measuring transactive memory systems in the field: scale development and validation.},
  author={Lewis, Kyle},
  journal={Journal of applied psychology},
  volume={88},
  number={4},
  pages={587},
  year={2003},
  publisher={American Psychological Association}
}

@article{peltokorpi2008transactive,
  title={Transactive memory systems},
  author={Peltokorpi, Vesa},
  journal={Review of general Psychology},
  volume={12},
  number={4},
  pages={378--394},
  year={2008},
  publisher={SAGE Publications Sage CA: Los Angeles, CA}
}

@article{van2025landscape,
  title={Landscape of AI literacy in education: approaches, impacts, and challenges for student preparedness—a narrative review},
  author={van der Linde, Guillermo and Rodriguez-Montoya, Cristobal and Garrido, Luis Eduardo},
  journal={Discover Education},
  volume={4},
  number={1},
  pages={561},
  year={2025},
  publisher={Springer}
}

@article{hui2024qwen2,
  title={Qwen2. 5-coder technical report},
  author={Hui, Binyuan and Yang, Jian and Cui, Zeyu and Yang, Jiaxi and Liu, Dayiheng and Zhang, Lei and Liu, Tianyu and Zhang, Jiajun and Yu, Bowen and Lu, Keming and others},
  journal={arXiv preprint arXiv:2409.12186},
  year={2024}
}

\end{document}